\documentclass[fleqn,10pt]{wlscirep}
\usepackage[utf8]{inputenc}
\usepackage[T1]{fontenc}

\title{The unintended consequences of inconsistent pandemic control policies}

\author[1,2,3,*]{Benjamin M.\ Althouse}
\author[4]{Brendan Wallace}
\author[5,6]{Brendan Case}
\author[7,8,9]{Samuel V.\ Scarpino}
\author[10,11]{Antoine Allard}
\author[12]{Andrew M.\ Berdahl}
\author[13,14]{Easton R.\ White}
\author[5,6,10]{Laurent H\'ebert-Dufresne}

\affil[1]{Institute for Disease Modeling, Global Health, Bill \& Melinda Gates Foundation, Seattle, WA}
\affil[2]{University of Washington, Seattle, WA 98105}
\affil[3]{New Mexico State University, Las Cruces, NM 88003}
\affil[4]{Department of Applied Mathematics, University of Washington, Seattle, WA 98195, USA}
\affil[5]{Department of Computer Science, University of Vermont, Burlington, VT 05405, USA}
\affil[6]{Vermont Complex Systems Center, University of Vermont, Burlington, VT 05405, USA}
\affil[7]{Network Science Institute, Northeastern University, Boston, MA, USA}
\affil[8]{Roux Institute, Northeastern University, Portland, ME, USA}
\affil[9]{Santa Fe Institute, Santa Fe, NM, USA}
\affil[10]{D\'epartement de physique, de g\'enie physique et d'optique, Universit\'e Laval, Qu\'ebec (Qu\'ebec), Canada G1V 0A6}
\affil[11]{Centre interdisciplinaire en mod\'elisation math\'ematique, Universit\'e Laval, Qu\'ebec (Qu\'ebec), Canada G1V 0A6}
\affil[12]{School of Aquatic \& Fishery Sciences, University of Washington, Seattle, WA 98195, USA}
\affil[13]{Department of Biology, University of Vermont, Burlington, VT 05405, USA}
\affil[14]{Gund Institute for Environment, University of Vermont, Burlington, VT 05405, USA}

\affil[*]{balthouse@idmod.org}

\makeatletter
\renewcommand{\@maketitle}{%
{%
\thispagestyle{empty}%
\vskip-36pt%
{\raggedright\sffamily\bfseries\fontsize{20}{25}\selectfont \@title\par}%
\vskip10pt
{\raggedright\sffamily\fontsize{12}{16}\selectfont  \@author\par}
\vskip25pt%
}%
}%
\makeatother

\begin{document}

\flushbottom
\maketitle

\\
\\
\noindent \textbf{Corresponding author:}\\
Benjamin M Althouse\\
Institute for Disease Modeling\\
Bill \& Melinda Gates Foundation\\
500 5th Ave N, Seattle, WA 98109\\
Phone: (425) 777-9615\\
Email: balthouse@idmod.org\\

\thispagestyle{empty}

\clearpage
\pagebreak

\textbf{
Controlling the spread of COVID-19 -- even after a licensed vaccine is available -- requires the effective use of non-pharmaceutical interventions, e.g., physical distancing, limits on group sizes, mask wearing, etc~\cite{la2020policy,tantrakarnapa2020influencing,lee2020interrupting,park2020information,cousins2020new,kraemer2020effect,anderson2020will}. To date, such interventions have not been uniformly and\slash or systematically implemented across the United States of America (US)~\cite{gupta2020tracking}. For example, even when under strict stay-at-home orders, numerous jurisdictions in the US granted exceptions and\slash or were in close proximity to locations with entirely different regulations in place. Here, we investigate the impact of such geographic inconsistencies in epidemic control policies by coupling high-resolution mobility, search, and COVID case data to a mathematical model of SARS-CoV-2 transmission. Our results show that while stay-at-home orders decrease contacts in most areas of the US, some specific activities and venues often see an increase in attendance. As an example, over the month of March 2020, between 10 and 30\% of churches in the US saw increases in attendance; even as the total number of visits to churches declined nationally. This heterogeneity, where certain venues see substantial increases in attendance while others close, suggests that closure can cause individuals to find an open venue, even if that requires longer-distance travel. And, indeed, the average distance travelled to churches in the US rose by 13\% over the same period, and over the summer, churches with more than 50 average weekly visitors saw an increase of 81\% in distance visitors had to travel to attend. Strikingly, our mathematical model reveals that, across a broad range of model parameters, partial measures can often be worse than no measures at all. In the most severe cases, individuals not complying with policies by traveling to neighboring jurisdictions can create epidemics when the outbreak would otherwise have been contained. Indeed, using county-level COVID-19 data, we show that mobility from high-incidence to low-incidence associated with travel for venues like churches, parks, and gyms consistently precedes rising case numbers in the low-incidence counties. Taken together, our data analysis of nearly 120 million church visitors across 184,677 churches, 14 million grocery visitors across 7,662 grocery stores, 13.5 million gym visitors across 5,483 gyms, 7.7 million cases across 3,195 counties, and modeling results highlight the potential unintended consequences of inconsistent epidemic control policies and stress the importance of balancing the societal needs of a population with the risk of an outbreak growing into a large epidemic, and the urgent need for centralized implementation and enforcement of non-pharmaceutical interventions.}

Severe acute respiratory syndrome coronavirus 2 (SARS-CoV-2, the virus that causes COVID-19) has swept the globe, revealing the strengths and weaknesses of our international, national, state, and local public health systems. Emerging evidence from countries such as Vietnam~\cite{la2020policy}, Thailand~\cite{tantrakarnapa2020influencing}, Singapore~\cite{lee2020interrupting}, South Korea~\cite{park2020information}, New Zealand~\cite{cousins2020new}, China~\cite{kraemer2020effect}, and others~\cite{anderson2020will} suggests that coordinated, national-level policies can control SARS-CoV-2 transmission. However, in many locations---in particular the United States of America---efforts to stem the spread of SARS-CoV-2 were instead implemented as a patchwork of self-isolation, school closures, and business restrictions~\cite{gupta2020tracking}. For example, throughout the months of March and April 2020, US states, counties, and cities often independently implemented stay-at-home orders, mask mandates, limits on gathering sizes, etc.~\cite{white2020state}. May and June saw nearly all states begin to reopen leading to increased cases through July, August, and September in turn leading towards closing again in half a dozen states including New York, California, and Texas~\cite{lee2020see}.
A potentially dire epidemiological consequence of this lack of coordination is that individuals can easily travel to areas with different control measures and avoid locally disrupted services and gatherings.

Even in countries with more uniform policies, religious activities have been the subject of much debate as the local risks associated with the the activity~\cite{pung2020investigation,yong2020connecting,james2020high} clashed with protections of the activity as an essential service to individuals and the community~\cite{del2020churchinaction}. Choirs and large services in particular have led to many superspreading events~\cite{SSEs}, with attack rates well-above 50\% in some cases~\cite{hamner2020high}. Unfortunately, there have been little efforts devoted to replacing religious services with safe alternatives, leading individuals to defy public health recommendations. As an example, individuals have defied church closures and attended mass gatherings, at times leading to legal prosecution~\cite{pastorArrested, pastorArrested2}. 

Other essential services have seen similar patterns, with public spaces such as urban and suburban parks and trails also being the subject of inconsistent visitation patterns and closures.
As other businesses close, there has been increased foot traffic in parks with many reporting overcrowding. When some, but not all, parks and trails close, individuals may travel further to areas remaining open potentially seeding virus to previously uninfected areas~\cite{parksOpen}.

Taken together, the non-uniform implementation and relaxation of US state-level interventions has left the country with high numbers of cases and potential distrust of public health interventions~\cite{GuideToGovenors}. Quantification of how movement patterns have changed from one's local closed business to a neighboring open business has been under-explored. Also needed are more general investigations into the impacts of non-uniform implementation of interventions on the transmission dynamics of SARS-CoV-2 and other pathogens. Here, we first examine online information seeking and physical foot traffic data to see how gathering-specific behavior has varied across the US during the COVID-19 pandemic. We then study a model of epidemics with partial gathering restrictions and partial adoption of said restrictions. Finally, we bridge the empirical data and model using county-level COVID case data showing that there is indeed an interaction between incidence and movement and demonstrate that COVID flows from counties with high incidence to counties with high incidence. We discuss the implications of these results especially as they relate to current discussions on relaxing or re-implementing stay-at-home orders and allowing gatherings.

\subsection*{Changing mobility and information-seeking}

We use data from SafeGraph to quantify human mobility after the adoption of physical distancing measures. SafeGraph is a data company that aggregates anonymized location data from numerous applications in order to provide insights about physical places. To enhance privacy, SafeGraph excludes census block group information if fewer than five devices visited an establishment in a month (two devices in a week) from a given census block group.
Using these data we use counts of visits and unique visitors to businesses across the US as well as the distance traveled from `home' (defined as the common nighttime location for the device over a 6 week period where nighttime is 6 pm to 7 am).

\begin{figure}
    \centering
    \includegraphics[width=\linewidth]{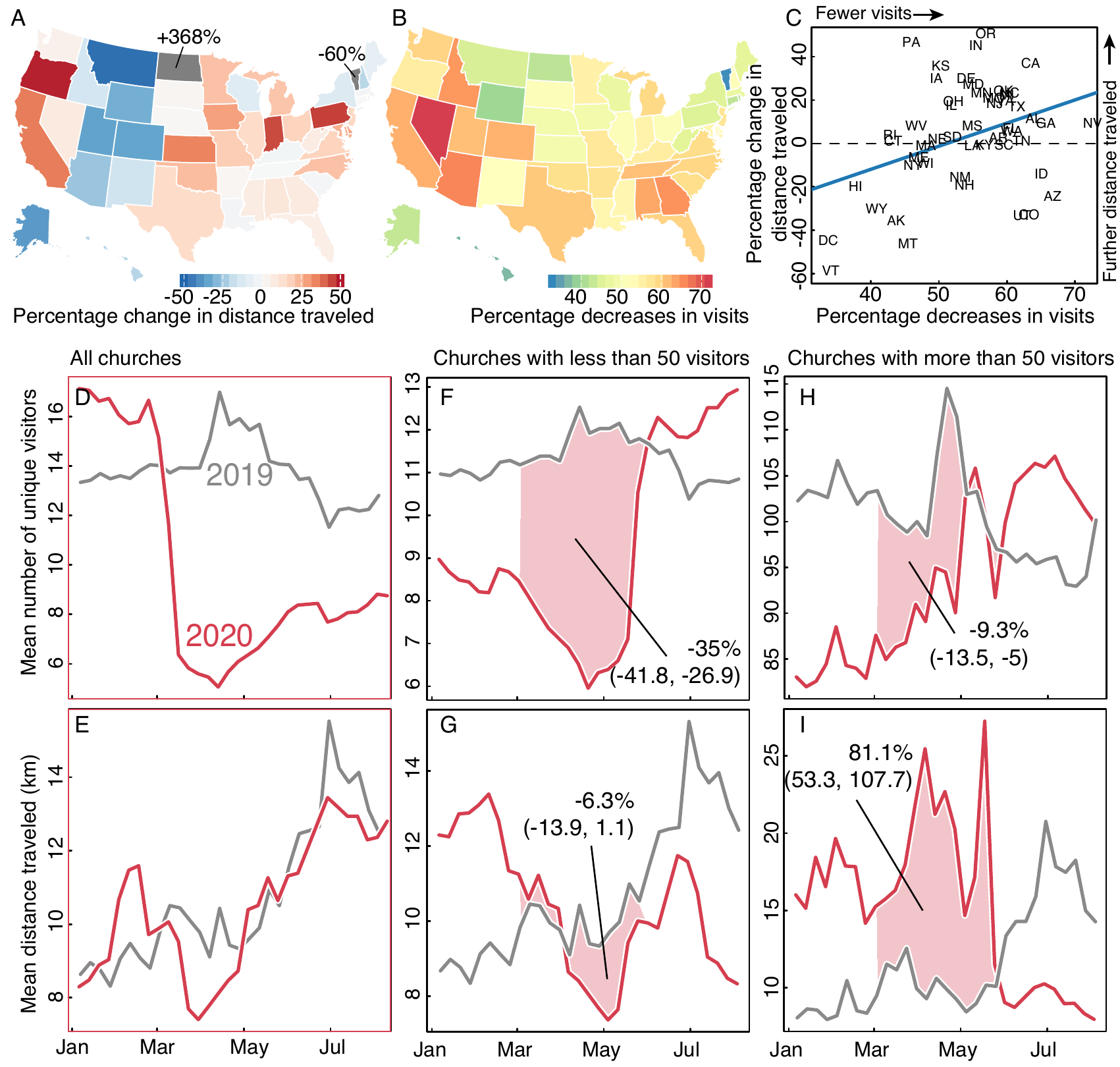}
    \caption{\textbf{Changes in distances traveled and number of visits to churches.} Panel A shows the percent change in distances traveled to churches by state over March 2020. Panel B shows the percent decreases in numbers of unique visitors to churches by state over March 2020. Panel C shows the relationship between percent changes in distance traveled by percent change in number of visits over March 2020. A decreasing number of visits, likely linked to restrictions and lockdown measures, is correlated with an increase in distance travelled. Panels D \& E show the mean unique visitors and distance traveled for all churches in the US from January through August comparing 2019 and 2020, respectively. Panels F \& G show visitors and distance in churches with less than 50 visitors, and panels H \& I show visitors and distance in churches with more than 50 visitors. While social distancing measures limited overall church visits, the distance traveled to large churches increased.
    }
    \label{fig:dists}
\end{figure}

\begin{figure}
    \centering
    \includegraphics[width=0.9\linewidth]{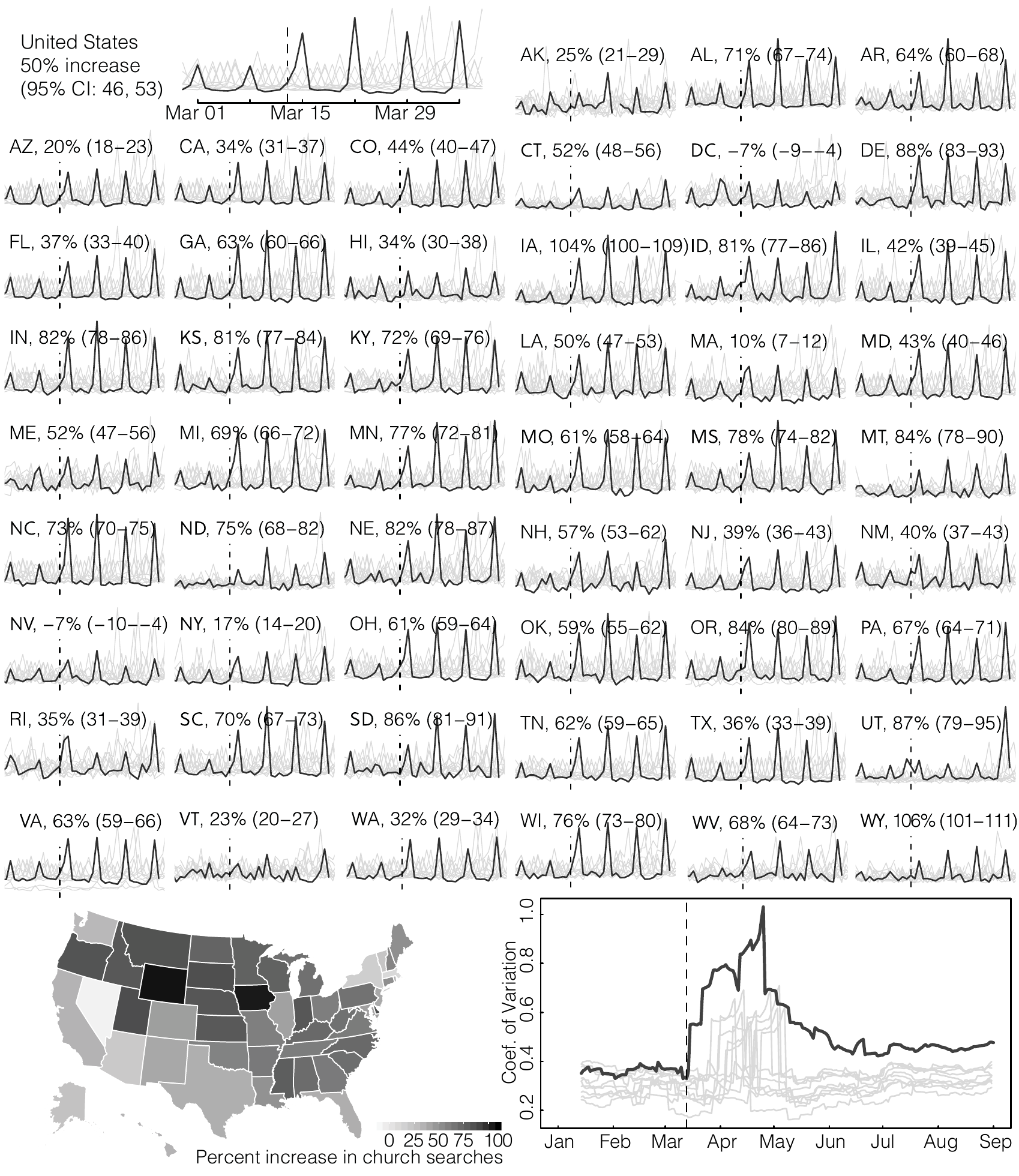}
    \caption{\textbf{Changes in information-seeking for churches and parks.} Sparklines show Google searches for ``church + churches'' (obtained using the Google Trends API for search) for all states in the US. Dark line indicates searches in 2020 and lighter lines 2010--2019.
    Percent increases are comparing Sunday search volume in 2020 to Sunday volumes in 2010-2019. 
    Map on bottom left shows the percent increases as displayed in the sparklines.
    Bottom right plot shows a 14-day running coefficient of variation (CV) of searches for ``church + churches''. 2020 saw significant increases in the CV that have remained elevated since Trump's declaration of a national emergency (March 13th, dashed vertical line).}
    \label{fig:search}
\end{figure}

What is clear from the data is that many individuals are willing to travel further to attend certain gatherings, in particular church services. We find that despite seeing an overall 56\% (95\% CI: 40--76) decrease in visits to churches comparing the first to the last week of March, individuals that do visit a church travel on average 13\% (95\% CI: 4--26) further to churches across most states in the country (Fig.~\ref{fig:dists}). The figure also shows that overall visits to churches remained low from March through August, the distance traveled to larger churches (those with more than a mean of 50 visitors per week) saw an 81.1\% (95\% CI: 53--108) increase in average distance traveled. 

That individuals are looking and traveling further for churches is also seen in Google search trends (as downloaded from the Google API for Trends), where queries for ``churches'' have increased since the beginning of March while searches for ``parks'' have decreased.  We compared search volumes for church on Sundays between March 13 and April 13, 2020 to Sundays in the previous 10 years of searches occurring on the same date across all US states. By normalizing to previous years, we are able to capture deviations during 2020 above and beyond typical searching patterns over this period, which encompasses Lent where individuals may have increased interest in attending church. Comparison to previous years should alleviate potential biases as the previous years act as counterfactuals to 2020. 
Overall, we see 49.6\% (95\% CI: 46.4--52.9) higher search volumes for churches in 2020 as compared to 2010--2019, with similarly higher volumes across states. Indeed, only Nevada and D.C. did not show statistically significantly higher search volumes in 2020, with the others ranging from 10\% (95\% CI: 7.0--11.8) in MA to 106\% (95\% CI: 100.5--111.3) in WY (Fig.~\ref{fig:search}).
Additionally, we compare a 14-day running coefficient of variation (CV) of church searches in 2020 to 2010--2019. We see an immediate spike in CV on the day Trump declared a national emergency, March 13th, which peaks in early May and remains elevated through August.
Conversely, we fit a series of sinusoidal regressions to estimate the expected number of searches for ``parks'' and find a 22\% (95\% CI, -10 to -31) decrease after the declaration of national emergency.
These patterns of searches indicate increased information-seeking for churches, potentially because an individual's normal church is closed. From a social network perspective, traveling to a more distant church is an unintended consequence where interventions might now be increasing contacts between communities.

Looking at the average attendance at churches, we find that while the number of visitors has decreased significantly overall, our result on increased travel were specifically driven by states with significant restrictions on visits and large decrease in visits. More specifically, we estimate an increase of about 5 additional km traveled for every 25\% decrease in visitors (95\% CI, 15--34, $p=0.002$). While we can not confirm the exact mechanism, these results are consistent with the idea that some individuals will travel further to seek an open church when their local church closes.

\begin{figure}
    \centering
    \includegraphics[width=\linewidth]{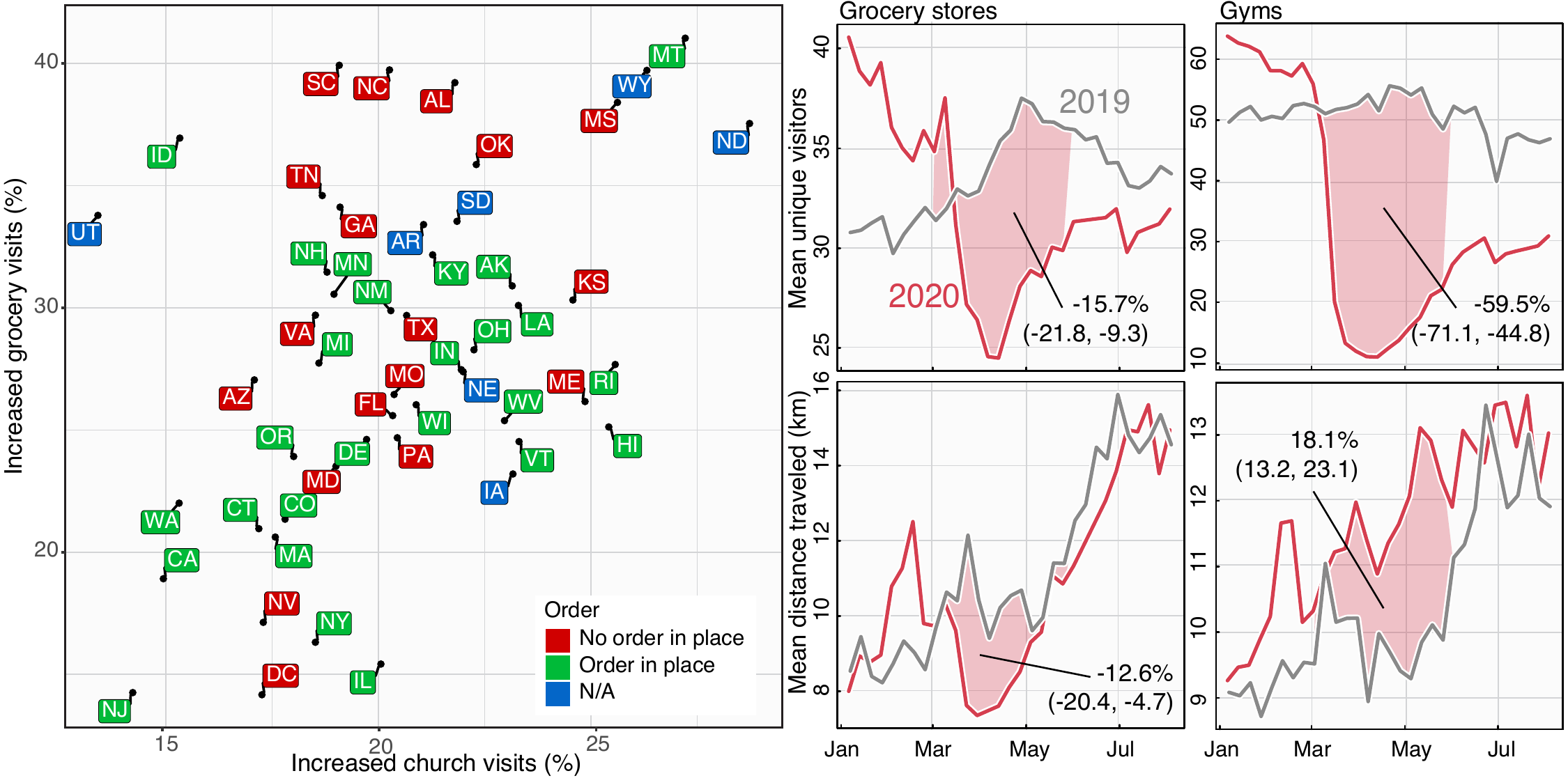}
    \caption{\textbf{Prevalence of churches and grocery stores with increased numbers of visitors and visitation patterns in 2019 and 2020.} Scatter plot of different states based on their increase in visits to essential services as well as whether the state had a stay-at-home order in place before March 29th. These are specific grocery stores or churches who had an increase in visits comparing the first and last week of March. The horizontal axis shows the fraction of churches with an increase in number of visits, and the vertical axis shows the fraction of grocery stores with an increase in number of visits. There is a positive correlation between both but not clear distinction with local policy, suggesting that the phenomenon is related to the local distribution of services and behavior of the local population. Right hand panels show the mean numbers of visitors for grocery stores and gyms (top panels) and the mean distance traveled (bottom panels). While visits and distance traveled both decreased for grocery stores -- as would be expected from movement restrictions in place -- most states saw increases in visits in 20\% to 40\% of grocery stores with the remainder driving the decrease. Visits to gyms also dropped drastically, but must like churches, with an associated and sizeable increase in the distance traveled.}
    \label{fig:increases}
\end{figure}

This phenomenon is not limited to religious services. We compared differences in numbers of visits to grocery stores (North American Industry Classification System [NAICS] code 4451) that had increases in visits to churches who had increases in visits. We find a correlation between increases to visits to churches with increases in visits to grocery stores (Pearson's r = 0.44) with increases to grocery stores being higher than to churches (slope = 0.9, 95\% CI: 0.37--1.44; Fig.~\ref{fig:increases}). Additionally we find increases to churches and grocery stores to be largely independent of whether the state had a stay-at-home order in place, suggesting that the phenomenon is closely related to the local distribution of services, individual burden such as food insecurity and behavior of the local population.
Comparing mean numbers of visitors and distance traveled for all grocery stores saw decreases in both -- as would be expected from movement restrictions in place -- gym visits on the other hand dropped drastically, but saw a sizeable increase in the distance traveled, which increased throughout the summer.

Even ignoring the fact that these additional visitors travel further and are therefore likely to increase the coupling between distinct communities, they also simply increase the number of contacts in their new church. Given that the expected number of contacts is expected to increase non-linearly with the number $n$ of participants in a gathering (i.e. potential contacts are proportional to $n(n-1)/2 \sim n^2$), it is unclear whether or not closing some churches might be worth the increased risk in the remaining open churches. To investigate this trade-off, we now design a simple model.

\subsection*{Mathematical model}

To more broadly explore the potential unintended consequences of inconsistent epidemic control policies, we formulated a simple, mathematical model which we call \textit{cloSIR} to couple disease dynamics with closure policies.  Specifically, we model an epidemic in a population of size $N$ uniformly distributed across $M$ gatherings of size $n = N/M$. We assume that a fraction $X$ of gatherings are closed at time $t_c$ to help contain an outbreak and that a fraction $Y$ of members in closed gatherings then decide to defy the closure by travelling to one of the remaining open gatherings. These open gatherings could be under a different set of rules in a different location or the venue\slash location itself may be defying government restrictions. Closures therefore protect the local community, which does comply with the closures, but can potentially increase attendance in any open venues.

As a first approximation, we ignore any spatial features and contacts occurring outside of average gatherings. We track Susceptible-Infectious-Recovered (SIR) dynamics within gatherings by assuming that the natural normalized transmission rate of the disease is $\lambda$ (with a recovery rate equal to 1 for time units set to the recovery period).
We use $S_{o/c}$, $I_{o/c}$ and $R_{o/c}$ to denote the number of susceptible, infectious and recovered individual in a typical open/closed gathering, respectively. Applying standard SIR dynamics in open gatherings but removing transmission events in closed gatherings, we write
\begin{align}
&\frac{dS_o}{dt} = -\lambda S_o I_o \qquad \frac{dI_o}{dt} = \lambda S_o I_o - I_o \qquad \dfrac{dR_o}{dt} = I_o \\
&\frac{dS_c}{dt} = 0 \qquad\qquad\;\;\; \frac{dI_c}{dt} = -I_c \qquad\qquad\;\;\; \frac{dR_c}{dt} = I_c \; .
\end{align}
The critical part of the cloSIR model is the implementation of closure policies at time $t_c$. At time $t<t_c$, all gatherings are open and we have $S_o+I_o+R_o = N/M$ and $S_c=I_c=R_c=0$ such that all derivatives are equal to zero in closed gatherings for $t<t_c$. Once the intervention is implemented at time $t=t_c$, we redistribute non-compliant individuals from closed to open gatherings.

Since $XM$ gatherings are closed, we have $XM\times(YN/M)=XYN$ non-compliant individuals to redistribute across $(1-X)M$ open gatherings. This connection between increases in intervention and displacement is based on the correlations observed in Fig.\ref{fig:dists}(c). This mechanism increases the population of open gatherings to $\frac{N}{M}\times[1+XY/(1-X)]$, and similarly decreases the population assigned to closed gatherings to $\frac{N}{M}(1-Y)$.

After closures are implemented, the dynamics of the cloSIR model are still governed by the same set of ordinary differential equations. As the outbreak progresses, we are interested in two key observables: first, the total number of infectious individuals
\begin{equation}
I(t) = 
  \begin{cases}
    MI_o(t), & \text{for } t<t_c \\
    (1-X)MI_o(t) + XMI_c(t), & \text{for } t\geq t_c
  \end{cases}
\end{equation}
and, second, the total fraction of recovered individuals
\begin{equation}
R(t) = 
  \begin{cases}
    MR_o(t), & \text{for } t<t_c \\
    (1-X)MR_o(t) + XMR_c(t), & \text{for } t\geq t_c.
  \end{cases}
\end{equation}
Finally, note that $N$ and $M$ are used to help us write the equations but only act as scale factors in our results and can therefore be set to $N=M=1$ for simplicity and without loss of generality.

\begin{figure}[ht]
    \centering
    \includegraphics[width=\linewidth]{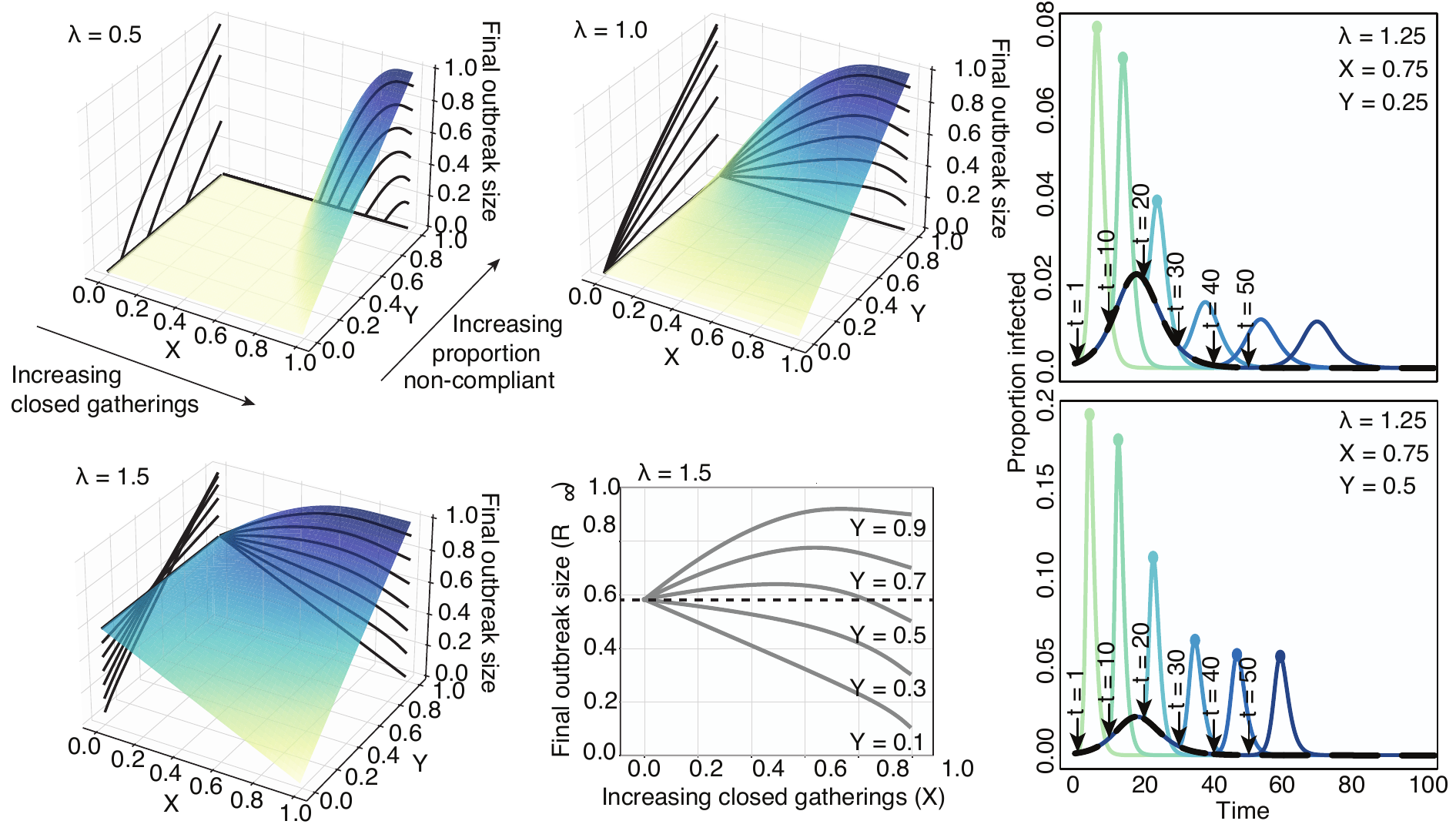}
    \caption{\textbf{Final state of the SIR dynamics with variable disease infectiousness and intervention scale.} Surfaces and contours show the final sizes of outbreaks across a range of intervention effects (X, proportion of closed gatherings) and proportion of non-compliant individuals who travel to open gatherings (Y). Three levels of infectiousness are illustrated ($\lambda = 0.5, 1.0, 1.5$). We see that the worst-case scenario is a function of $\lambda$, $X$ and $Y$, as there is no consistent ranking in outbreak sizes. Small interventions appear beneficial against very transmissible pathogens but risk lowering the epidemic threshold at high frequency of non-compliant individuals, and larger interventions accentuate this effect. 
    The bottom middle panel highlights this effect: we see that the impact of the fraction of non-compliant individuals is non-linear close to the epidemic threshold. At low values of $Y$, i.e., in a population with high compliance to recommendations, closing more gatherings is always beneficial. At the opposite end, for high values of $Y$, keeping all gatherings open is the optimal intervention.  However, at medium values of $Y$, while closing all gatherings is still the optimal intervention, keeping all gatherings open is better than partial closures. 
     The left-hand panels show the effects of changing intervention time across ranges of Y. The black curve depicts the course of the outbreak without any intervention. The various colored curves peeling off from the black curve show the course of the outbreak given differently timed interventions.
    Colored dots indicate epidemic peaks larger than the no intervention baseline scenario.
    Intuitively, we find that earlier interventions are always better and that delayed and imperfect interventions can cause second epidemic waves.
  }
    \label{fig:model1}
\end{figure}

Ultimately, although the dynamics are governed by the standard SIR differential equations for all time, the cloSIR model offers an interesting trade-off between controlling transmission by closing venues and \textit{intensifying} transmission by aggregating contacts in a smaller number of still open venues. The question then becomes whether the redistribution of participants among gathering locations, e.g., churches or parks, will have a positive or negative impact on the epidemic. Assuming one cannot ensure the closure of all venues, is closing a certain percentage of venues worth the increase in visitors to those that remain open?

Strikingly, we find that in many scenarios the optimal strategy to minimize the size of the outbreak is often no intervention at all.
Figure~\ref{fig:model1} shows that depending on the proportion of the population that chooses to go to another open business ($Y$) the final outbreak size is often minimized when $X=0$ (no closures).
In fact, close or below the epidemic threshold $\lambda_c = 1$, interventions can spark an outbreak in communities otherwise not at risk by increasing the concentration of susceptible individuals.
However, for stronger epidemics (larger $\lambda$), although a complete closure of gatherings $X=1$ might be the optimal strategy, the expected outbreak size often follows a non-monotonous function of $X$ such that the optimal outcome at $X=1$ is next to a worst-case scenario at large values of $X$ just below 1. What this implies is that the outcome is highly dependent on the amount of non-compliance that one can expect in a population (i.e. larger values of non-compliance $Y$).

Similarly, poorly-timed interventions can actually lead to additional waves of infection.
Figure~\ref{fig:model1} shows that secondary peaks of infection occur if intervention is initiated too late. Interestingly, stronger interventions tend to dramatically heighten the epidemic peak under many closure scenarios (colored curves) compared to the no-closure baseline (black curve).

The cloSIR model therefore provides a very simple and telling illustration of the potential impact of the collective behaviour observed in the empirical mobility and search data from the US around essential services. Although future research should layer in additional complexity into models of policy interventions, within the idealized scenario considered by the model, one can solve for specific features: E.g., the final outbreak size, the optimal closure percentage $X$, and the critical value of non-compliance $Y$ such that weak interventions increase outbreak size. Analyses of these different questions are presented in the Supplementary Information document.

\subsection*{Bridging movement, modeling, and incidence}

\begin{figure}[ht!!!!!!!!!!!!!!!]
    \centering
    \includegraphics[width=0.9\linewidth]{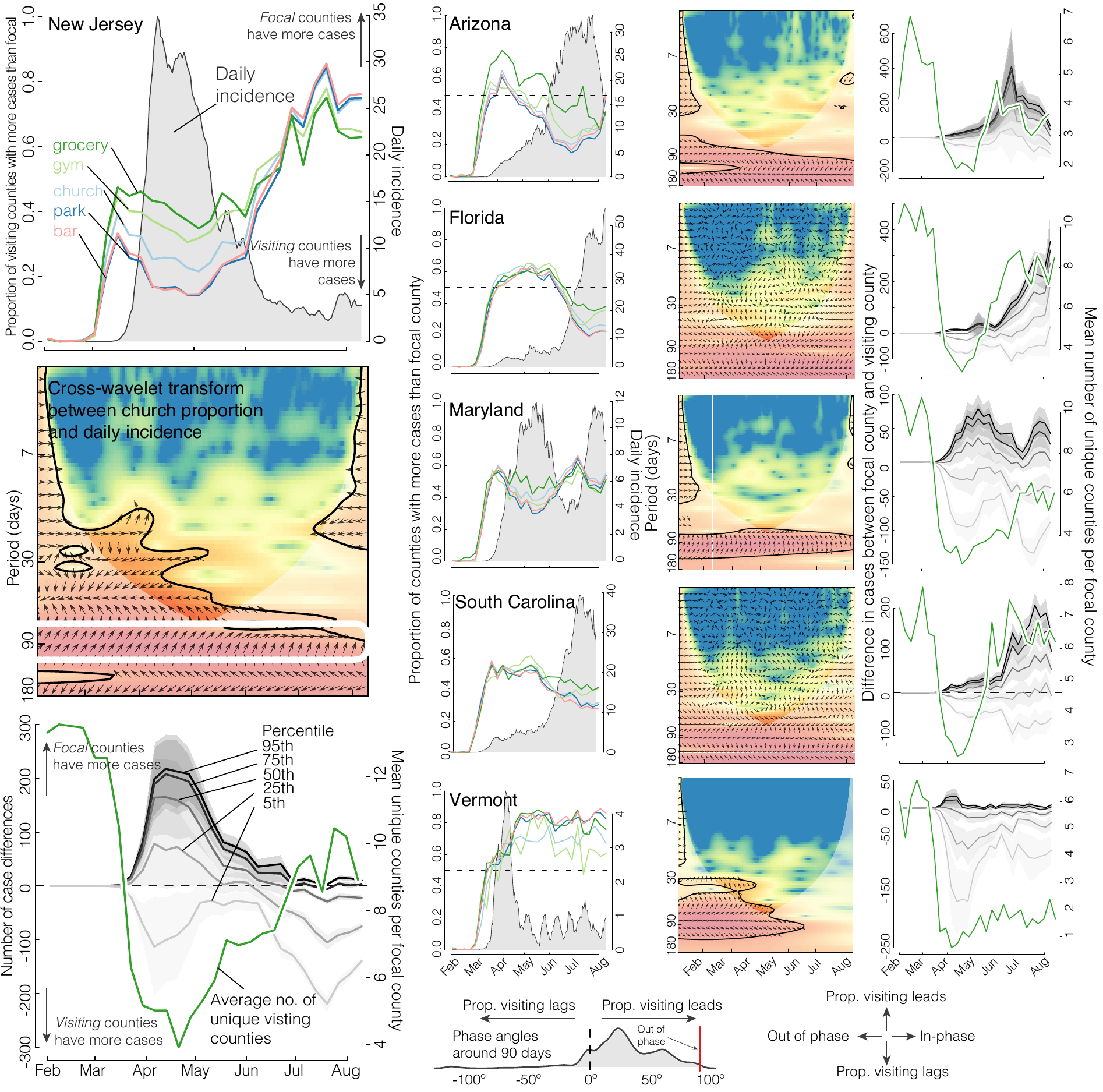}
    \caption{\textbf{Population movement is influenced by COVID incidence and is highly heterogeneous across states.} Left-hand column shows summaries of cases, movement, and county opening and closing in New Jersey, and serves as a guide to interpreting the other panels. New Jersey saw a surge in incidence in early April (7-day running mean, grey shade, top panel). We distinguish between the {\em focal county}, which is the county receiving visitors from listed {\em visiting counties}, which are recorded in the SafeGraph data set. We can then calculate the  proportion of visiting counties which have more cases than the focal county being visited for churches, gyms, grocery stores, parks, and bars (colored lines in the top left panel). When the proportion of visiting counties have more cases than the focal county being visited we see a rise in incidence, i.e., a net flux of cases in to a county from high incidence counties causes an overall increase in cases. As the proportion shifts from visiting counties having more cases to focal counties, overall cases are low or decline. These phenomena are seen across the 50 US States and DC (bottom phase angle density). To more formally quantify this, we computed cross-wavelet transforms~\cite{althouse2018seasonality,grinsted2004application,cazelles2008wavelet} to assess whether the proportion of visiting cases is in- or out of phase with incidence. Colors indicate power of the transform at a given time (x-axis) and periodicity (y-axis, in days), thick black lines denote statistically significant interactions, and arrows indicate the phase angle with in-phase pointing right, out-of-phase pointing left, and proportion of visiting counties larger than focal leading incidence by 90$^{\circ}$ pointing straight up.
    Strong coherence is seen around 90 days with proportion of visiting cases leading overall incidence across the US (circled in white). Finally, we look at the magnitude of population flux in response to incidence in the bottom panels. Grey lines are the 95th, 75th, 50th, 25th, and 5th quantiles of the difference in the numbers of cases between the focal county and the visiting county. There is substantial variation in the numbers of population movement across states. The green line indicates the average number of unique visiting counties per focal county for that state. While most states saw a rebound in the number of unique visiting counties, Vermont has maintained a low number of influx of population. Plots for all 50 states and D.C. are presented as supplementary materials.
    }
    \label{fig:case.data}
\end{figure}

Finally, the natural question suggested by the mobility data and the cloSIR model is: {\em Does differential mobility from non-uniform policy implementations lead to unintended consequences in incidence?} SARS-CoV-2 incidence data were downloaded from the COVID-19 Data Repository by the Center for Systems Science and Engineering at Johns Hopkins University at the county level beginning in February, 2020~\cite{dong2020interactive}. Figure~\ref{fig:case.data} summarizes the results. We first distinguish between a {\em focal county}, which is the county receiving visitors from other, listed {\em visiting counties}, which are recorded in the SafeGraph data set. We can then calculate the proportion of visiting counties which have more cases than the focal county being visited for churches, gyms, grocery stores, parks, and bars. Nearly uniformly we see that when the visiting counties have more cases than the focal county, COVID incidence goes up, and it goes down when the focal county has more cases. This rise or fall in the relative sizes of focal and visiting counties precedes rising incidence by about 90 days as evidenced through cross-wavelet analyses~\cite{althouse2018seasonality,grinsted2004application,cazelles2008wavelet}. Indeed, 85.5\% of phase angles around 90 days are over zero indicating a lead for proportion of sizes over incidence. We also find large heterogeneity in the magnitude of population movement and cases across states (see Supplemental figures), as well as in the average number of unique counties visiting a focal county across states. For example, while all states saw a large decline in unique visiting counties, states such as Florida and South Carolina saw rapid rebounds to pre-closure levels (around two months) leading to an increase in incidence, contrasted to Vermont which has kept unique visiting counties low and subsequently has not seen a rise in cases (Figure~\ref{fig:case.data}). Taken together, these results indicate an interaction between SARS-CoV-2 transmission and population movement.

\section*{Discussion}

Using real-time mobility and search data in the US, we found that while overall visits to various types of venues decreased in response to state-level lock-downs, the average distance traveled increased significantly, indicating that individuals are traveling further to attend these venues. This is corroborated by Google search queries for ``churches'' which increased across the month of March, indicating an increase in information seeking for churches in general. Through a mathematical model -- the cloSIR model -- we found under various circumstances some intervention can be worse than none at all. Finally, using county-level COVID case data we found that incidence influenced individual movement and vise-versa, where visiting counties having higher cases than focal counties increase incidence at a lag of 90 days. These source-sink dynamics were further modified by the average number of unique visiting counties where states that saw a rapid return to normal unique visiting counties saw large increases in incidence (such as in South Carolina and Florida) and states that maintained a limited number of unique visiting counties saw no increases in cases (Vermont). Taken together, these results indicate the importance of uniform implementation of non-pharmaceutical interventions (NPIs) with centralized leadership and enforcement, ideally at the national level.

Balancing the mental, economic, and social health of populations with the serious risks of COVID-19 means that the decision to implement movement restrictions (e.g., \emph{cordons sanitaires}) should be carefully considered and if chosen implemented uniformly. This is the same for other NPIs (e.g., hand washing, facial covering, social distancing, etc). Indeed, evidence from China suggests that while the cordon sanitaire of Wuhan delayed the outbreak, it was local measures that slowed transmission and ultimately controlled it~\cite{pan2020association,kraemer2020effect}. Despite evidence of the efficacy of consistent NPIs, many countries (especially the United States of America) continue to implement control measures in a scattered, patchwork manner~\cite{gupta2020tracking}. 

While distance traveled and the number of visits to essential services did not correlate strongly with any demographic variables (e.g. population density, average age), both of these responses did correlate with community tightness, with tight communities being those ``with strong norms and little tolerance for deviance" (Fig. S1). Gelfand et al. (2020) found that countries with both efficient governments and those with tight cultures were the most effective in limiting COVID-19 cases and deaths~\cite{gelfand2020importance}. 
However, White \& H\'ebert-Dufresne (2020) found the opposite for the US, with tighter states having faster COVID-19 growth rates early in the pandemic\cite{white2020state}. 
In the context of our cloSIR model, if a government expects compliance issues and complete lockdown is not possible, it might be best to have no lockdown at all (Figure~\ref{fig:model1}). 
This is an extreme example and we do not advocate foregoing NPIs, but uniform implementation and enforcement is critical.
Indeed, if one is in a county with high incidence movement restrictions should be put in place to limit travel to lower incidence counties (Figure~\ref{fig:case.data}).

Our results are particularly relevant as the loosening and re-tightening of restrictions on businesses and other institutions has been occurring in a disorganized and spatially heterogeneous way across the US~\cite{covidamp}. As we saw during the inconsistent shutdown, this lack of coordination at the policy-level creates incentives for individuals to travel to those areas which are open to purchase products and services, to access public areas, like beaches and parks, and\slash or attend religious services. For instance, the partial reopening of businesses in Georgia on Apr 24th, led to a 13\% increase in visitors from nearby states~\cite{UMcovid}. This surge in inter-state mobility both increases distance traveled and the amount of clustering in a limited number of areas. The same result occurs if individual businesses or institutions decide on different reopening strategies. For instance, in Burlington, Vermont retail stores were allowed to open with limited capacity on May 18th. While many stores opened on this date, many store owners decided to postpone openings out of safety concerns~\cite{VTreopen}. 
Thus, if a state relaxes restrictions on businesses, but some businesses owners choose to remain closed, this has the same effect demonstrated in our analyses. 
Scattered or disorganized reopening after lock-down can therefore spark new waves of infection. 
Critically, reopening more slowly is not necessarily better if done non-uniformly.

There are several caveats to our study. First, we developed a simple model that was able to illustrate the potential unintended consequences of individuals adapting their behavior to seek essential services under inconsistent physical distancing policies. While the simplicity of this model is a strength when trying to isolate the effects of inconsistent control policies on COVID-19 transmission, future work will be needed before such models could be used to actively inform specific policy decisions. Second, because the SafeGraph data do not track individual users over long periods of time, those observed in late March are not necessarily the same individuals observed earlier in the month. Moreover, we may expect biases in the diversity and behaviors of individuals tracked by the system since different types of gatherings attract different individuals.
Altogether, these limitations mean that small geographic regions should not be directly compared to one another, or even to themselves at a different time, and different locations should not be directly compared. 
This is why we coarse-grained our results over states, why we mostly compared relative changes and not absolute differences, and why we attempted to correlate our findings with a secondary data source like online searches. 
Future work is therefore warranted, on both data collection and analysis (comparing changing movement patterns for various other business types) and mathematical modeling (expansion to include more metapopulation structure to explore the interplay of interventions across scales).

As we have shown, it is of key importance that NPIs for the fight against COVID-19--- specifically related to business and venue closure---be implemented in a uniform way. Similarly, relaxation of such interventions must be done methodically and over time, with a strong emphasis on equity across incomes and geographies to avoid endangering individuals with lower socioeconomic status~\cite{weill2020social}. The difficulty of this situation is vast and it requires strong, centralized implementation and management that would be best coordinated at a national level.  
Human behavior is a strong driver of the transmission dynamics of SARS-CoV-2 and care must be taken to reduce the heavy burden imposed by COVID-19 and avoid unintended, negative consequences from inconsistent policies around implementing and relaxing NPIs.

\section*{Acknowledgements}

We thank Edward Wenger and Josh Proctor for helpful comments on the manuscript. B.C. is supported as a Fellow of the National Science Foundation under NRT award DGE-1735316.  S.V.S. is supported by startup funds provided by Northeastern University. E.R.W. was supported by the COVID-19 Rapid Research Fund from the Gund Institute for Environment at the University of Vermont. L.H.-D. acknowledges support from the National Institutes of Health 1P20 GM125498-01 Centers of Biomedical Research Excellence Award. The funders had no role in study design, data collection, data analysis, the decision to publish, or preparation of the manuscript. 

The content of this article is solely the responsibility of the authors and does not necessarily represent the official views of their respective employers or funders.

\section*{Author contributions}
B.M.A. conceived the study, designed experiments, performed statistical calculations of the movement, search, and case data, and wrote the first draft of the manuscript. B.W., B.C., A.M.B., E.R.W., and L.H-D. conceived and coded the cloSIR model. L.H-D. and S.V.S. contributed to study design and revision of the manuscript. All authors contributed intellectually to the study and revision of the manuscript.

\section*{Competing interests}

All authors declare no competing interests exist.

\section*{Data availability statement}

All data are freely available from their respective sources.

\section*{Code availability statement}

Code for the cloSIR model is fully available online at https://github.com/LaurentHebert/cloSIR.

\clearpage
\pagebreak

\bibliographystyle{ieeetr}
\bibliography{R0}

\end{document}